\documentclass[aps,prb,amsmath,showpacs,twocolumn]{revtex4}

\usepackage{graphicx}

	\newcommand{\textmarkright}{December 4, 2006\ (submitted to Appl.\ Phys.\ Lett.)}
\begin{document} 
\title{Hysteretic ac loss of superconducting strips simultaneously exposed to ac transport current and phase-different ac magnetic field}
\author{Yasunori Mawatari}
\affiliation{%
	National Institute of Advanced Industrial Science and Technology (AIST), \\
	Tsukuba, Ibaraki 305--8568, Japan
}
\author{Kazuhiro Kajikawa}
\affiliation{%
	Research Institute of Superconductor Science and Systems, 
	Kyushu University, \\
	6--10--1 Hakozaki, Higashi-ku, Fukuoka 812--8581, Japan
}
\received{December 4, 2006}

\begin{abstract}
A simple analytical expression is presented for hysteretic ac loss $Q$ of a superconducting strip simultaneously exposed to an ac transport current $I_0\cos\omega t$ and a phase-different ac magnetic field $H_0\cos(\omega t+\theta_0)$. 
On the basis of Bean's critical state model, we calculate $Q$ for small current amplitude $I_0\ll I_c$, for small magnetic field amplitude $H_0\ll I_c/2\pi a$, and for arbitrary phase difference $\theta_0$, where $I_c$ is the critical current and $2a$ is the width of the strip. 
The resulting expression for $Q=Q(I_0,H_0,\theta_0)$ is a simple biquadratic function of both $I_0$ and $H_0$, and $Q$ becomes maximum (minimum) when $\theta_0=0$ or $\pi$ ($\theta_0=\pi/2$). 
\end{abstract}
\maketitle
	\thispagestyle{myheadings}\markright{\textmarkright}

Hysteretic alternating current (ac) loss is one of the most important parameters of superconducting wires for electrical power devices. 
In three-phase ac cables, for example, electrical wires are simultaneously subjected to ac transport currents and to phase-different ac magnetic fields. 

High-temperature superconducting wires have strip geometry, and theoretical expressions for hysteretic ac losses for a superconducting strip have been derived by Norris~\cite{Norris70} for ac transport currents and by Halse~\cite{Halse70} and Brandt {\em et al}.~\cite{Brandt93a} for ac magnetic fields, based on the critical state model.~\cite{Bean62} 
Behavior of a superconducting strip exposed to both a transport current and an applied magnetic field is complicated,~\cite{Brandt93b,Zeldov94,Schonborg01,Pardo} and the hysteretic ac loss of a strip simultaneously subjected to ac transport current and ac magnetic field with arbitrary phase difference has not yet been analytically investigated. 
Here, we derive an analytical expression of the hysteretic ac loss of a superconducting strip simultaneously exposed to an ac transport current and a phase-different ac magnetic field. 

The superconducting strip that we consider has infinite length along the $z$ axis, and has a flat rectangular cross section in which $|x|<a$ and $|y|<d/2$, where $2a\gg d$. 
For simplicity, we consider the thin-strip limit of $d/2a\to 0$. 
The behavior of such a thin strip is described by a perpendicular magnetic field $H_y(x)$ at $y=0$ and a sheet current $K_z(x)$. 
In the strip, we assume the critical current density $j_c$ to be uniform and constant as in the Bean model,~\cite{Bean62} and the critical current is given by $I_c=2j_cad$. 

The transport current $I_t$ flows along the $z$ direction (i.e., longitudinal direction) in the strip, and the magnetic field $H_a$ is applied along the $y$ direction (i.e., direction perpendicular to the width of the strip). 
Both $I_t(t)$ and $H_a(t)$ are sinusoidal functions of time $t$ with identical angular frequency $\omega$, and are given by 
\begin{eqnarray}
	I_t(t) &=& I_0\cos\omega t , 
\label{It}\\
	H_a(t) &=& H_0\cos(\omega t+\theta_0) , 
\label{Ha}
\end{eqnarray}
where $I_0$ is the current amplitude, $H_0$ is the magnetic field amplitude, and $\theta_0$ is the phase difference. 
The hysteretic ac loss $Q=Q(I_0,H_0,\theta_0)$ of a strip per unit length per ac cycle is independent of $\omega$, and is a function of $I_0$, $H_0$, and $\theta_0$. 
To derive a simple analytical expression for $Q$, we confine our theoretical calculation to small ac amplitudes, $I_0\ll I_c$ and $H_0\ll I_c/2\pi a=j_cd/\pi$.

First, we consider a strip carrying a dc transport current $I_t=I_0$ that is monotonically increased from zero. 
For $I_0\ll I_c$, the current and magnetic field distributions near the edges at $x=\pm a$ play crucial roles. 
In the ideal Meissner state we have~\cite{Brandt93b,Zeldov94} $H_y(x)=0$ and $K_z(x)=(I_0/\pi)/\sqrt{a^2-x^2}$ for $|x|<a$, and $H_y(x)=(I_0/2\pi)\mbox{sgn}(x)/\sqrt{x^2-a^2}$ for $|x|>a$. 
The $H_y(x)$ and $K_z(x)$ near the edge at $x\simeq\pm a$ are reduced to 
\begin{equation}
	H_y(x)\simeq \frac{\varphi_{\pm}}{\sqrt{|x|-a}} , \quad
	K_z(x)\simeq \frac{\pm 2\varphi_{\pm}}{\sqrt{a-|x|}} , 
\label{HyKz_edge-Meissner}
\end{equation}
where 
\begin{equation}
	\varphi_{\pm}=\pm I_0/2\pi\sqrt{2a} . 
\label{phi_I0}
\end{equation}
In the critical state model~\cite{Bean62} the $H_y(x)$ and $K_z(x)$ near the edge at $x=+a$ should satisfy $H_y(x)=0$ for $x<\alpha_0$ and $K_z(x)=j_cd$ for $\alpha_0<x<a$, where $\alpha_0$ is the parameter for the flux front. 
The corresponding expressions for $x\simeq a\simeq\alpha_0$ are~\cite{Norris70,Brandt93b,Zeldov94} 
\begin{eqnarray}
	H_y(x) &\simeq& 
	\begin{cases}
		0 & \mbox{for $x<\alpha_0$} , \\
	\displaystyle
		\frac{j_cd}{\pi} \,\mbox{arctanh}
		\left(\sqrt{\frac{x-\alpha_0}{a-\alpha_0}}\right) 
		& \mbox{for $\alpha_0<x<a$} , \\[2ex]
	\displaystyle
		\frac{j_cd}{\pi} \,\mbox{arctanh}
		\left(\sqrt{\frac{a-\alpha_0}{x-\alpha_0}}\right) 
		& \mbox{for $x>a$} . 
	\end{cases}
\nonumber\\
\label{Hy_edge-critical}\\
	K_z(x) &\simeq& 
	\begin{cases}
	\displaystyle
		\frac{2j_cd}{\pi} \,\arctan
		\left(\sqrt{\frac{a-\alpha_0}{\alpha_0-x}}\right) 
		& \mbox{for $x<\alpha_0$} , \\
		\ j_cd & \mbox{for $\alpha_0<x<a$}
	\end{cases}
\nonumber\\
\label{Kz_edge-critical}
\end{eqnarray}
Equation~\eqref{Hy_edge-critical} for $x>a$ with $\alpha_0\to a$ is reduced to $H_y(x)\to (j_cd/\pi) \sqrt{(a-\alpha_0)/(x-a)}$, which must coincide with $H_y(x)$ in Eq.~\eqref{HyKz_edge-Meissner}. 
The parameter $\alpha_0$ is, therefore, determined by 
\begin{equation}
	a-\alpha_0 =(\pi|\varphi_+|/j_cd)^2 . 
\label{alpha-phi}
\end{equation}

When a superconducting strip carries an ac transport current given by Eq.~\eqref{It}, the hysteretic ac loss arising from the edge at $x=+a$ is calculated from $H_y(x)$ for dc current by using Eqs.~\eqref{Hy_edge-critical} and \eqref{alpha-phi}: 
\begin{eqnarray}
	Q_+ &=& 4\mu_0j_cd \int_{\alpha_0}^a dx (a-x)H_y(x) 
\label{Q+_Hy}\\
	&\simeq& (4/3\pi)\mu_0 (j_cd)^2 (a-\alpha_0)^2 
\label{Q+_alpha}\\
	&\simeq& (4\pi^3/3)\mu_0 |\varphi_+|^4/(j_cd)^2 . 
\label{Q+_phi}
\end{eqnarray}
Calculation of $Q_-\propto|\varphi_-|^4$ arising from the edge at $x=-a$ is similar to that of $Q_+\propto|\varphi_+|^4$, and the total loss $Q=Q_+ +Q_-$ is given by 
\begin{equation}
	Q \simeq \frac{4}{3}\pi^3 \mu_0 
		\frac{|\varphi_+|^4 +|\varphi_-|^4}{(j_cd)^2} . 
\label{Q-total_phi}
\end{equation}
As seen from Eqs.~\eqref{HyKz_edge-Meissner} and \eqref{Q-total_phi}, 
the ac loss for small ac amplitude is directly related to the field 
distributions in the ideal Meissner state.~\cite{Kajikawa05,Mawatari06} 
Substitution of Eq.~\eqref{phi_I0} into Eq.~\eqref{Q-total_phi} yields 
\begin{equation}
	Q\simeq (\mu_0 I_c^2/6\pi)\, j_0^4 ,  
\label{Q_j0-Norris}
\end{equation}
where $j_0$ is the reduced current amplitude, 
\begin{equation}
	j_0 =I_0/I_c =I_0/2j_cad . 
\label{j0}
\end{equation}
Equation~\eqref{Q_j0-Norris} corresponds to the theoretical result derived by Norris~\cite{Norris70} for $j_0\ll 1$. 

\begin{figure*}[bth]
\includegraphics{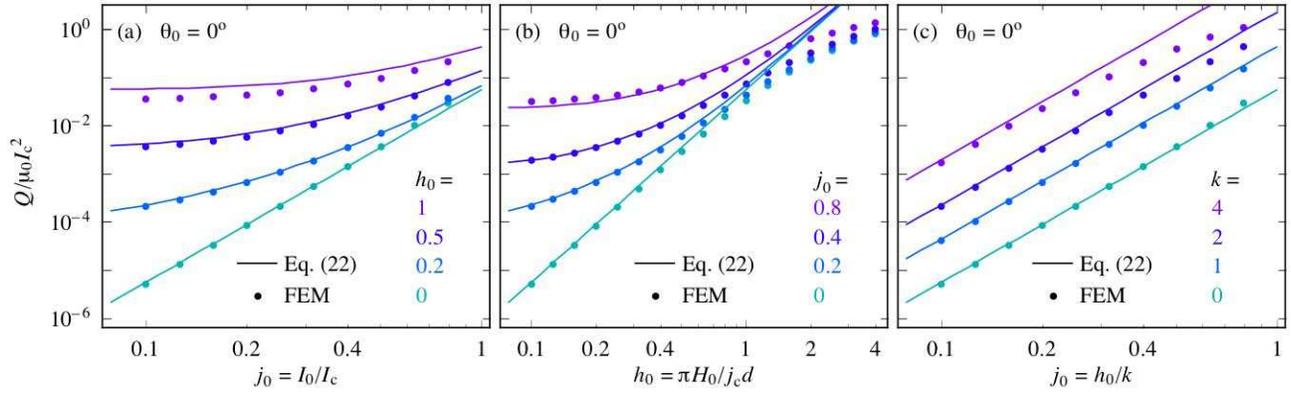}
\caption{%
Double-logarithmic plots of the hysteretic ac loss $Q/\mu_0I_c^2$ for $\theta_0=0$, as a function of the current amplitude $j_0=I_0/I_c$ and the magnetic field amplitude $h_0=\pi H_0/j_cd= 2\pi aH_0/I_c$: 
(a) $Q$ vs.\ $j_0$ for $h_0=0,\,0.2,\,0.5,$ and $1$, 
(b) $Q$ vs.\ $h_0$ for $j_0=0,\,0.2,\,0.4,$ and $0.8$, 
and (c) $Q$ vs.\ $j_0=h_0/k$ for $k=0,\,1,\,2,$ and 4. 
The lines show the analytical results calculated using Eq.~\eqref{Q_j0-h0-p0}, and the symbols show the numerical results calculated using FEM. 
}
\label{Fig_Q-jh}
\end{figure*}

Next we consider a superconducting strip exposed to a magnetic field $H_a=H_0$, which is monotonically increased from zero. 
For $H_0\ll I_c/2\pi a=j_cd/\pi$, the current and magnetic field distributions near the edges at $x=\pm a$ play crucial roles. 
In the ideal Meissner state we have~\cite{Brandt93a,Brandt93b,Zeldov94} $H_y(x)=0$ and $K_z(x)=2H_0 x/\sqrt{a^2-x^2}$ for $|x|<a$, and $H_y(x)=H_0 |x|/\sqrt{x^2-a^2}$ for $|x|>a$. 
The $H_y(x)$ and $K_z(x)$ near the edge at $x\simeq\pm a$ are given by Eq.~\eqref{HyKz_edge-Meissner}, where $\varphi_{\pm}$ is 
\begin{equation}
	\varphi_{\pm}= H_0\sqrt{a/2} . 
\label{phi_H0}
\end{equation}
Equations~\eqref{Hy_edge-critical}, \eqref{Kz_edge-critical}, and \eqref{alpha-phi} are valid also for a strip exposed to a magnetic field. 
When a superconducting strip is exposed to an ac magnetic field given by Eq.~\eqref{Ha}, the hysteretic ac loss is also calculated by substituting Eq.~\eqref{phi_H0} into Eq.~\eqref{Q-total_phi}. 
The resulting ac loss of a strip in an ac magnetic field is given by 
\begin{equation}
	Q\simeq (\mu_0 I_c^2/6\pi)\, h_0^4 ,  
\label{Q_h0-Halse}
\end{equation}
where $h_0\ll 1$ is the reduced field amplitude defined by 
\begin{equation}
	h_0=2\pi aH_0/I_c= \pi H_0/j_cd . 
\label{h0}
\end{equation}
Equation~\eqref{Q_h0-Halse} corresponds to the theoretical result derived by Halse~\cite{Halse70} for $h_0\ll 1$. 

Now, let us consider a superconducting strip simultaneously exposed to an ac transport current given by Eq.~\eqref{It} and an ac magnetic field given by Eq.~\eqref{Ha}. 
In the ideal Meissner state, $K_z(x,t)$ for $|x|<a$ and $H_y(x,t)$ for $|x|>a$ are given by~\cite{Brandt93b,Zeldov94} 
\begin{eqnarray}
	K_z(x,t) &=& \frac{2}{\sqrt{a^2-x^2}} 
		\left[ \frac{I_t(t)}{2\pi} +H_a(t)x \right] , 
\label{Kz_Meissner}\\
	H_y(x,t) &=& \frac{\mbox{sgn}(x)}{\sqrt{x^2-a^2}} 
		\left[ \frac{I_t(t)}{2\pi} +H_a(t)x \right] , 
\label{H_y_Meissner}
\end{eqnarray}
respectively. 
Substitution of Eqs.~\eqref{It} and \eqref{Ha} into Eq.~\eqref{Kz_Meissner} yields approximate expressions for $K_z$ and $H_y$ near the edges of a strip $x\simeq\pm a$, as 
\begin{eqnarray}
	K_z(x,t) &\simeq& \frac{\pm 2\varphi_{\pm}}{\sqrt{a-|x|}} 
		\cos(\omega t+\vartheta_{\pm}) , 
\label{Kz_Meissner-edge}\\
	H_y(x,t) &\simeq& \frac{\varphi_{\pm}}{\sqrt{|x|-a}} 
		\cos(\omega t+\vartheta_{\pm}) , 
\label{Hy_Meissner-edge}
\end{eqnarray}
where $|\varphi_{\pm}|$ is given by 
\begin{equation}
	|\varphi_{\pm}|= \frac{1}{2\pi\sqrt{2a}} 
		\left[ I_0^2 +(2\pi aH_0)^2 
		\pm 4\pi aI_0H_0\cos\theta_0 \right]^{1/2} . 
\label{phi_I0-H0}
\end{equation}
Because $\vartheta_{\pm}$ in Eqs.~\eqref{Kz_Meissner-edge} and \eqref{Hy_Meissner-edge} do not appear in the following calculations, we do not show the details here for $\vartheta_{\pm}$. 
The hysteretic ac loss of a superconducting strip simultaneously exposed to an ac transport current given by Eq.~\eqref{It} and an ac magnetic field given by Eq.~\eqref{Ha} is obtained by substituting Eq.~\eqref{phi_I0-H0} into Eq.~\eqref{Q-total_phi}. 
The resulting expression for the hysteretic ac loss of a superconducting strip per unit length per cycle is given by 
\begin{equation}
	Q\simeq \frac{\mu_0 I_c^2}{6\pi} 
		\Bigl[ j_0^4 +h_0^4 
		+2j_0^2h_0^2\left(1+2\cos^2\theta_0\right) \Bigr] . 
\label{Q_j0-h0-p0}
\end{equation}
This simple expression is the main result of the present paper. 
We see that Eq.~\eqref{Q_j0-h0-p0} is the generalization of Eq.~\eqref{Q_j0-Norris} for $h_0=0$ and Eq.~\eqref{Q_h0-Halse} for $j_0=0$. 

Equation~\eqref{Q_j0-h0-p0} has been derived assuming that the critical sheet-current density $j_cd$ is uniform and is independent of $x$ (i.e., strips with uniform $j_c$ and with rectangular cross section). 
When $j_cd$ is nonuniform, however, the ac loss is generally given by 
\begin{eqnarray}
	Q &\propto& |\varphi_+|^m +|\varphi_-|^m 
\nonumber\\
	&\propto& \left( j_0^2 +h_0^2 +2j_0h_0\cos\theta_0 \right)^{m/2} 
\nonumber\\
	&& {}+ \left( j_0^2 +h_0^2 -2j_0h_0\cos\theta_0 \right)^{m/2} , 
\label{Q_j0-h0-p0-m}
\end{eqnarray}
where the parameter $m$ depends on the behavior of $j_cd$ near the edges of the strips. 
For example, $m=4$ [Eq.~\eqref{Q_j0-h0-p0}] for constant $j_cd$, $m=3$ for $j_cd\propto (1-|x|/a)^{1/2}$ (e.g., strips with uniform $j_c$ and with elliptic cross-section), and $m= 4(p+1)/(2p+1)$ for $j_cd\propto (1-|x|/a)^p$.~\cite{Kajikawa04} 
Equation~\eqref{Q_j0-h0-p0-m} with $m=3$ is similar to the ac loss of superconducting slabs~\cite{Carr79,Ashworth00} and solenoids.~\cite{Kawasaki01}

Figure~\ref{Fig_Q-jh} shows $Q$ for $\theta_0=0$ as a function of $j_0$ and $h_0$: (a) $Q$ vs.\ $j_0$ for fixed $h_0$, (b) $Q$ vs.\ $h_0$ for fixed $j_0$, and (c) $Q$ vs.\ $j_0$ with $h_0=kj_0$ for fixed $k=h_0/j_0$.~\cite{Pardo} 
The lines show analytical results calculated using Eq.~\eqref{Q_j0-h0-p0}, and the symbols show numerical results calculated by using a finite-element method (FEM) to solve Maxwell equations.~\cite{Kajikawa06} 
As seen in Fig.~\ref{Fig_Q-jh}, the range of $(j_0,h_0)$ in which Eq.~\eqref{Q_j0-h0-p0} is valid is not restricted to the small-ac limit, $j_0\ll 1$ and $h_0\ll 1$. 
Comparison of the analytical result from Eq.~\eqref{Q_j0-h0-p0} and the numerical results shown in Fig.~\ref{Fig_Q-jh} confirms that the relative error of Eq.~\eqref{Q_j0-h0-p0} for $\theta_0=0$ is less than 10\% when $j_0<0.4$ and $h_0<0.4$.

\begin{figure}[b]
\includegraphics{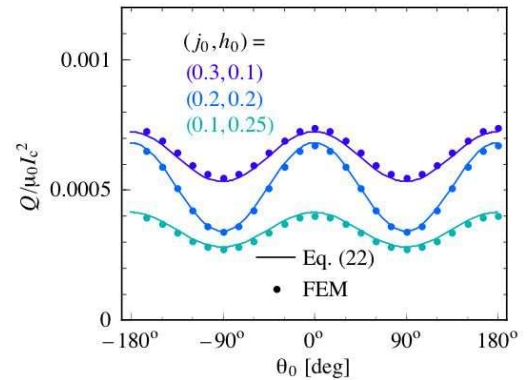}
\caption{%
Hysteretic ac loss $Q/\mu_0I_c^2$ vs.\ phase difference $\theta_0$ for $(j_0,h_0)=(0.3,0.1), (0.2,0.2),$ and $(0.1,0.25)$. 
The lines show the analytical results calculated using Eq.~\eqref{Q_j0-h0-p0}, and the symbols show the numerical results calculated using FEM. 
}
\label{Fig_Q-th}
\end{figure}

Figure~\ref{Fig_Q-th} shows $Q$ vs.\ $\theta_0$ for fixed $j_0$ and $h_0$. 
The data from the FEM numerical calculation agrees well with Eq.~\eqref{Q_j0-h0-p0} for small $(j_0,h_0)$ as shown in Fig.~\ref{Fig_Q-th}. 
As actually observed in Refs.~\onlinecite{Ashworth00} and \onlinecite{Ashworth99}, the theoretical $Q$ given by Eqs.~\eqref{Q_j0-h0-p0} and \eqref{Q_j0-h0-p0-m} is maximum when $\theta_0=0$ or $\pi$, and is minimum when $\theta_0=\pi/2$. 
Note that the experimental $Q$ can be maximum (minimum) when $\theta_0\neq 0$ ($\theta_0\neq \pi/2$), as reported in Refs.~\onlinecite{Nguyen05,Gomory06,Vojenciak06}. 
The reason for the shifts in $\theta_0$ for maximum and minimum $Q$ is that the ac magnetic field was large (i.e., $h_0>1$) in those measurements.~\cite{Nguyen05,Gomory06,Vojenciak06}

In summary, we theoretically investigated the hysteretic ac loss of a superconducting strip simultaneously exposed to an ac transport current [Eq.~\eqref{It}] and an ac magnetic field [Eq.~\eqref{Ha}]. 
When $j_cd$ is uniform, the ac loss of a strip of unit length for one ac cycle is given by Eq.~\eqref{Q_j0-h0-p0}, where $j_0$ and $h_0$ are defined by Eqs.~\eqref{j0} and \eqref{h0}, respectively. 
When $j_cd$ is nonuniform near the edges of a strip, on the other hand, the ac loss is proportional to the right-hand side of Eq.~\eqref{Q_j0-h0-p0-m}. 
The simple analytical result of Eq.~\eqref{Q_j0-h0-p0} was derived here assuming small ac amplitudes, and we confirmed that the relative error in Eq.~\eqref{Q_j0-h0-p0} is less than $10\%$ when $j_0<0.4$ and $h_0<0.4$.

We thank M.\ Furuse, K.\ Develos-Bagarinao, and H.\ Yamasaki 
for stimulating discussions.


\begin{thebibliography}{99} 
\bibitem{Norris70}
W. T. Norris, J. Phys. D {\bf 3}, 489 (1970).
\bibitem{Halse70}
M. R. Halse, J. Phys. D {\bf 3}, 717 (1970).
\bibitem{Brandt93a}
E.H. Brandt, M.V. Indenbom, and A. Forkl, 
Europhys Lett. {\bf 22}, 735 (1993). 
\bibitem{Bean62}
C. P. Bean, \prl {\bf 8}, 250 (1962); \rmp {\bf 36}, 31 (1964).
\bibitem{Brandt93b}
E.H. Brandt and M. Indenbom, 
\prb {\bf 48}, 12 893 (1993). 
\bibitem{Zeldov94}
E. Zeldov, J. R. Clem, M. McElfresh, and M. Darwin, 
\prb {\bf 49}, 9802 (1994).
\bibitem{Schonborg01}
N. Sch\"{o}nborg, J. Appl. Phys. {\bf 90}, 2930 (2001).
\bibitem{Pardo}
E. Pardo, F. G\"{o}m\"{o}ry, J. \v{S}ouc, and J. M. Ceballos, 
cond-mat/0510314.
\bibitem{Kajikawa05}
K. Kajikawa, T. Hayashi, and K. Funaki, 
Cryogenics {\bf 45}, 289 (2005).
\bibitem{Mawatari06}
Y. Mawatari and K. Kajikawa, \apl {\bf 88}, 092503 (2006).
\bibitem{Kajikawa04}
K. Kajikawa, Y. Mawatari, T. Hayashi, and K. Funaki, 
Supercond. Sci. Technol. {\bf 17}, 555 (2004).
\bibitem{Carr79}
W. J. Carr, IEEE Trans. Magn. {\bf MAG-15}, 240 (1979).
\bibitem{Ashworth00}
S. P. Ashworth and M. Suenaga, 
Physica C {\bf 329}, 149 (2000).
\bibitem{Kawasaki01}
K. Kawasaki, K. Kajikawa, M. Iwakuma, and K. Funaki, 
Physica C {\bf 357-360}, 1205 (2001).
\bibitem{Kajikawa06}
K. Kajikawa, Y. Mawatari, Y. Iiyama, T. Hayashi, K. Enpuku, K. Funaki, 
M. Furuse, and S. Fuchino,
Physica C {\bf 445-448}, 1058 (2006).
\bibitem{Ashworth99}
S. P. Ashworth and M. Suenaga, 
Physica C {\bf 315}, 79 (1999).
\bibitem{Nguyen05}
D. N. Nguyen, P. V. P. S. S. Sastry, D. C. Knoll, G. Zhang, and J. Schwartz, 
J. Appl. Phys. {\bf 98}, 073902 (2005).
\bibitem{Gomory06}
F. G\"{o}m\"{o}ry, J. \v{S}ouc, M. Vojen\v{c}iak, E. Seiler, B. Klin\v{c}ok, J. M. Ceballos, E. Pardo, A. Sanchez, C. Navau, S. Farinon, and P. Fabbricatore, 
Supercond. Sci. Technol. {\bf 19}, S60 (2006).
\bibitem{Vojenciak06}
M. Vojen\v{c}iak, J. \v{S}ouc, J. M. Ceballos, F. G\"{o}m\"{o}ry, B. Klin\v{c}ok, E. Pardo, and F. Grilli, 
Supercond. Sci. Technol. {\bf 19}, 397 (2006).
\end{thebibliography}
\end{document}